\documentclass[hyper]{JHEP} 

\usepackage{epsfig}

\newcommand\fverb{\setbox\pippobox=\hbox\bgroup\verb}

\newcommand\fverbdo{\egroup\medskip\noindent%

            \fbox{\unhbox\pippobox}\ }

\newcommand\fverbit{\egroup\item[\fbox{\unhbox\pippobox}]}

\newbox\pippobox

\title{Canonical Analysis of Non-Relativistic String
with Non-Relativistic World-Sheet}
\author{J. Kluso\v{n}\\
Department of
Theoretical Physics and Astrophysics\\
Faculty of Science, Masaryk University\\
Kotl\'{a}\v{r}sk\'{a} 2, 611 37, Brno\\
Czech Republic\\
E-mail: \email{klu@physics.muni.cz}} \preprint{}

 \abstract{We perform canonical analysis of non-relativistic
 string theory with non-relativistic world-sheet gravity. We determine
structure of constraints and symplectic structure of canonical variables.}

\def\tpi{\tilde{\pi}}

\def\be{\begin{equation}}

\def\ee{\end{equation}}

\def\bea{\begin{eqnarray}}

\def\eea{\end{eqnarray}}

\def\tmH{\tilde{\mH}}

\def\mH{\mathcal{H}}

\newcommand{\mF}{\mathcal{F}}

\newcommand{\mG}{\mathcal{G}}

\newcommand{\bT}{\mathbf{T}}

\newcommand{\mL}{\mathcal{L}}

\def\pb #1{\left\{#1\right\}}

\begin{document}
\section{Introduction and Summary}
AdS/CFT correspondence is the most known example of 
holographic duality \cite{Maldacena:1997re}. This correspondence,
in its strongest form, claims that $SU(N)$ $\mathcal{N}=4$ SYM theory in four dimensions  is equivalent to type IIB theory on $AdS_5\times S^5$ at any values
of $N$ and   'tHooft coupling $\lambda$. On the other hand understanding
this duality at the strongest form is still lacking and hence we should restrict
to some limits of this correspondence. 

Recently such an interesting limit was suggested in 	\cite{Harmark:2014mpa}
and it is known as Spin Matrix Theory (SMT) and describes near BPS limit of AdS/CFT.
It is quantum mechanical theory with Hamiltonian given as sum of harmonic oscillator operators that transform both in adjoin representation of $SU(N)$ and in a particular spin subgroup $G_s$ of the global superconformal $PSU(2,2|4)$ symmetries of $\mathcal{N}=4$.

One can ask the question what is the dual description of this quantum mechanical model. It was suggested \cite{Harmark:2017rpg} and further studied in \cite{Harmark:2018cdl,Harmark:2019upf,Harmark:2020vll}
that dual theory in the bulk corresponds to non-relativistic string theory with 
non-relativistic world-sheet known as SMT string. These special non-relativistic theories should be considered
in the broader context of non-relativistic string theories that were studied 
recently in \cite{Harmark:2017rpg,Harmark:2019upf,Harmark:2018cdl,Harmark:2020vll} and also 
\cite{Andringa:2012uz,Gomis:2020izd,Gomis:2020fui,Kluson:2020aoq,Hansen:2020pqs,Yan:2019xsf,Kluson:2019xuo,Bergshoeff:2019pij,Gallegos:2019icg,Gomis:2019zyu,Kluson:2019ifd,Kluson:2018vfd,Kluson:2018grx,Bergshoeff:2018yvt,Kluson:2018egd}. This development is related to the generalization of  Newton-Cartan geometry \cite{Cartan:1923zea}
to the stringy Newton-Cartan geometry
\cite{Andringa:2012uz} and torsional Newton-Cartan geometry. Moreover, SMT string  was derived in \cite{Harmark:2017rpg,Harmark:2019upf,Harmark:2018cdl} by specific non-relativistic limit on the world-sheet of non-relativistic string in torsional NC background. Recently this SMT string was very intensively studied in 
\cite{Harmark:2020vll} where particular class of backgrounds for SMT string, known as flat-fluxed backgrounds,  was analysed. 
 In these backgrounds SMT string
reduces to a free theory. These world-sheet theories are analogues of the Polyakov action on Minkowski target space-time. 

The next step would be to analyse properties of SMT string in general background. In order to do this we should certainly study classical dynamics as for example its Hamiltonian form. The aim of this paper is to find such a formulation in the most general case. 

Let us be more explicit. We start with the action for SMT string that was found in 
\cite{Harmark:2018cdl} and perform canonical analysis of this theory. As opposite to Polyakov form of the relativistic string now the action is formulated using vierbein $e_\alpha^{ \ a}$ where $\alpha=0,1$ correspond to world-sheet coordinates while $a=0,1$ correspond  to tangent space coordinates. Note that  $e_\alpha^{ \ a}$ is invertible matrix with inverse $\theta^\alpha_{ \ a}$. Now it is crucial that 
the quadratic term with $\partial_\alpha x^\mu\partial_\beta x^\nu h_{\mu\nu}$ is multiplied with $\theta^\alpha_{ \ 1}\theta^\beta_{ \ 1}$ as opposite to the relativistic case when this term has the form $\theta^\alpha_{ \ a}\theta^\beta_{ \ b}\eta^{ab}$. Then it is necessary to distinguish two cases. In the first case we presume that $\theta^0_{ \ 1}\neq 0$. Then the relation between momenta and time derivative of $x^\mu$ is invertible. As a result we obtain Hamiltonian together with set of the primary constraints that follow from the structure of the theory. Careful analysis of the preservation of the primary constraints gives two secondary constraints that are first class constraints that reflect the fact that the theory is invariant under world-sheet diffeomorphism. We also identify four additional second class constraints and Poisson brackets between them. Finally we determine symplectic structure for canonical variables which is given in terms of the Dirac brackets. We identify that in this case the Dirac brackets coincide with Poisson brackets. 

The situation is different when $\theta^0_{ \ 1}=0$. In this case it is not possible
to express time derivative of $x^\mu$ using canonical variables. Instead we get new $d-$constraints where $d-$is number of dimensions labelled with $x^\mu$. Then the canonical analysis
is slightly more complicated than in previous case. However we again find two first class constraints that reflect invariance of the action under reparameterization. We further identify second class constraints and Poisson brackets between them. The presence of these constraints then imply non-trivial symplectic structure 
between canonical variables $x^\mu$ which confirms analysis presented in 
\cite{Harmark:2020vll}.

Let us outline our results and suggest further directions of research. We found Hamiltonian formulation  of SMT string and we identified structure of constraints. We discussed two cases when in the first one we were able to invert relation between time derivative of $x^\mu$ and canonical momenta. In fact, this is the most general
situation where all components of $\theta^\alpha_{ \ 1}$ are non-zero. On the other hand the second case when $\theta^0_{ \ 1}=0$ deserves separate treatment. This fact suggests that the spatial gauge as was used in \cite{Harmark:2020vll} cannot be reached from the general Hamiltonian. It is instructive to compare this situation with the standard relativistic Lagrangian where the relation between momenta
and $\partial x^\mu$ contains expression $\theta^0_{ \ a}\eta^\alpha_{ \ b}\eta^{ab}
\partial_\alpha x^\mu$ that can be certainly inverted even if we impose condition 
$\theta^0_{ \ 1}=0$. On the other hand when we studied the situation when $\theta^0_{ \ 1}=0$ separately we found theory with non-trivial symplectic structure as in 
\cite{Harmark:2020vll}.

Certainly this work can be extended in many directions. It would be nice to study the most general form of the string with the non-relativistic world-sheet 
and study its consistency from canonical point of view. It would be also extremely interesting to study supersymmetric generalization of this two dimensional theory. 

This paper is organized as follows. In the next section (\ref{second}) we review basic properties of non-relativistic string and we perform canonical analysis it the most general case. We also determine symplectic structure of given theory. In section (\ref{third}) we separately discuss the case $\theta^0_{\ 1}=0$ and we determine corresponding Hamiltonian and symplectic structure.

\section{Hamiltonian Analysis of SMT String }\label{second}
We begin with the Polyakov form of the action for SMT string   that was introduced in \cite{Harmark:2018cdl} and that has the form
\begin{equation}\label{actnonnon}
S=-\frac{T}{2}\int d^2\sigma 
(2 \epsilon^{\alpha\beta}m_\alpha\partial_\beta \eta+e\theta^\alpha_{ \ 1}\theta^\beta_{ \ 1}h_{\alpha\beta}+
\omega \epsilon^{\alpha\beta}e_\alpha^{ \ 0}\tau_\beta+
\psi \epsilon^{\alpha\beta}(e_\alpha^{ \ 0}\partial_\beta \eta+e_\alpha^{ \ 1}\tau_\beta)) \ . 
\end{equation}
Let us explain meaning of various symbols that appear in (\ref{actnonnon}). The world-sheet is labelled by $\sigma^0,\sigma^1\equiv\sigma$ and  $T$ is string tension. Further, $m_\mu,h_{\mu\nu}$ and $\tau_\mu$ are target space-time Newton-Cartan fields that obey conditions
\begin{equation}\label{defNC}
\tau_\mu h^{\mu\nu}=0 \ , \quad  v^\mu h_{\mu\nu}=0 \ , \quad \tau_\mu v^\mu=-1 \ , \quad  h_{\mu\nu}h^{\nu\rho}
-\tau_\mu v^\rho=\delta_\mu^\rho \ . 
\end{equation}
The world-sheet metric is defined with the help of zwiebein $e_\alpha^{ \ a} \ , a=0,1$  with inverse $\theta^\alpha_{ \ a}$ that obey
\begin{equation}
e_\alpha^{ \ a}\theta^\alpha_{\ b}=\delta^a_b \ , \quad 
e_\alpha^{ \ a}\theta^\beta_{ \ a}=\delta_\alpha^\beta \ .
\end{equation}
As was argued in \cite{Harmark:2018cdl} the world-sheet theory is non-relativistic since $e_\alpha^{ \ a}$ play different role in the action. This can be already seen 
from (\ref{actnonnon}) since zweibein inverse $\theta^\alpha_{ \ a}$ does not appear
in Lorentz invariant way $\theta^\alpha_{ \ a}\theta^\beta_{ \ b}\eta^{ab}$ but instead there is an expression $
\theta^\alpha_{ \ 1}\theta^\beta_{ \ 1}$. This fact has an important consequence
for the structure of this theory. Note also that 
\begin{equation}
e=\det e_\alpha^{ \ a}
\end{equation}
and 
\begin{equation}
m_\alpha=m_\mu\partial_\alpha x^\mu \ , \quad 
h_{\alpha\beta}=h_{\mu\nu}\partial_\alpha x^\mu \partial_\beta x^\nu \ ,  \quad 
\tau_\alpha=\tau_\mu\partial_\alpha x^\mu \ ,
\end{equation}
where $x^\mu$ label embedding of the string into target space-time. Finally 
$\eta$ is scalar field defined on world-sheet. 

We should stress that the theory is manifestly invariant under world-sheet
diffeomorphism $\sigma'^\alpha=f^\alpha(\sigma)$ where world-volume fields
transform as
\begin{equation}
x'^\mu(\sigma')=x^\mu(\sigma) \ , \quad 
\eta'(\sigma')=\eta(\sigma) \ , \quad 
e_\beta^{'  \ b}(\sigma')=e^b_{ \ \alpha}(\sigma)\frac{\partial \sigma^\alpha}{
\partial \sigma'^\beta} \ . 
\end{equation}
Our goal is to find Hamiltonian formulation of this theory in order to investigate possible non-relativistic nature of it. First of all we start with 
the definition of conjugate momenta. From (\ref{actnonnon}) we obtain 
\begin{eqnarray}\label{defmom}
& &\pi^\alpha_{ \ b}=\frac{\partial \mL}{\partial (\partial_0 e_\alpha^{ \ a})}\approx 0 
\ , \quad p_\psi=\frac{\partial \mL}{\partial (\partial_0 \psi)} \approx 0 \ , \quad  \pi_{\omega}=\frac{\partial \mL}{\partial (\partial_0\omega)}\approx 0 \ ,
\nonumber \\
& & 
p_\eta=\frac{\partial\mL}{\partial (\partial_0\eta)}=T m_1+\frac{T}{2}\psi e_1^{ \ 0} \ , 
\nonumber \\
& & p_\mu=\frac{\partial \mL}{\partial (\partial_0 x^\mu)}=
-Tm_\mu\partial_1 \eta-T e \theta^0_{ \ 1}\theta^\beta_{ \ 1}
h_{\mu\nu}\partial_\beta x^\nu+\frac{T}{2}\omega e_1^{ \ 0}\tau_\mu+\frac{T}{2}\psi e_1^{ \ 1}\tau_\mu \ . \nonumber \\
\nonumber \\
\end{eqnarray}
It is clear that definition of $p_\eta$ implies following primary constraint
\begin{eqnarray}
\Sigma_1\equiv p_\eta-T m_1-\frac{T}{2}\psi e_1^{ \ 0}\approx 0  \ .
\nonumber \\
\end{eqnarray}
In this section we will presume that $\theta^0_{ \ 1}$ is non-zero
and hence we can express time derivative of $x^\mu$ as function of  $p_\mu$. On the other hand 
there is another 
primary constraint that follows 
 from the definition of $p_\mu$ given in (\ref{defmom})
\begin{eqnarray}\nonumber \\
\Sigma_2\equiv
v^\mu p_\mu+Tv^\mu m_\mu \partial_1 \eta+
\frac{T}{2}\omega e_1^{ \ 0}+\frac{T}{2}\psi e_1^{ \ 1} 
\approx 0 \ 
\end{eqnarray}
using $v^\mu h_{\mu\nu}=0 \ , \quad v^\mu\tau_\mu=-1$. 

Returning to  (\ref{defmom}) we obtain bare Hamiltonian density in the form
\begin{eqnarray}
& &\mH_B=p_\mu\partial_0x^\mu+p_\eta\partial_0\eta-\mL=
\nonumber \\
& &=-\frac{1}{2Te \theta^0_{ \ 1}\theta^0_{ \ 1}}
[p_\mu h^{\mu\nu}p_\nu+2Tp_\mu h^{\mu\nu}m_\nu\partial_1\eta+T^2\partial_1\eta
m_\mu h^{\mu\nu}m_\nu\partial_1\eta]-\nonumber \\
& &-\frac{\theta^0_{ \ 1}\theta^1_{ \ 1}}{\theta^0_{ \ 1}\theta^0_{ \ 1}}(p_\mu \partial_1 x^\mu+p_\mu v^\mu \tau_1+\partial_1\eta m_1+
\partial_1 \eta m_\mu v^\mu \tau_1)
+\frac{T}{2}\omega e_0^{ \ 0}\tau_1 +
\frac{T}{2}\psi (e_0^{ \ 1}\tau_1+
e_0^{ \ 0}\partial_1\eta) \ . 
\nonumber \\
\end{eqnarray}
As is well known from the theory of systems with constraints 
the time evolution is governed by extended Hamiltonian that incorporates
bare Hamiltonian together with set of all primary constraints. Explicitly
we have
\begin{eqnarray}
\mH_E=
\mH_B+
\Omega^1 \Sigma_1+\Omega^2 \Sigma_2+\Omega_\alpha^{ \ a}\pi^\alpha_{ \ a}+\Omega_\psi p_\psi+
\Omega_\omega p_\omega \ , 
\nonumber \\
\end{eqnarray}
where $\Omega^1,\Omega^2,\Omega_\alpha^{ \ a},\Omega_\psi$ and $\Omega_\omega$
are Lagrange multipliers.

Now we should analyse condition of the preservation of all primary constraints
$\pi^\alpha_{ \ a}\approx 0,p_\omega\approx 0 \ , p_\psi\approx 0,
\Sigma_1\approx 0 \ , \Sigma_2\approx 0$. To do this we need following
canonical Poisson brackets 
\begin{eqnarray}
& &\pb{e_\alpha^{ \ a}(\sigma),\pi^\beta_{ \ b}(\sigma')}=
\delta_\alpha^\beta \delta_b^a \delta(\sigma-\sigma') \ , \nonumber \\
& &\pb{\psi(\sigma),p_\psi(\sigma')}=\delta(\sigma-\sigma') \ , \quad 
\pb{\omega(\sigma),p_\omega(\sigma')}=\delta(\sigma-\sigma') \ . \nonumber \\
\end{eqnarray}
First of all we have that $\Sigma_{1,2}$ are second class constraints together
with
$p_\psi,p_\omega$ 
as follows from   Poisson brackets
\begin{eqnarray}
& &\pb{p_\psi(\sigma),\Sigma_1(\sigma')}=\frac{T}{2}e_1^{ \ 0}(\sigma)\delta(\sigma-\sigma') \ , \quad 
\pb{\pi^1_{ \ 0}(\sigma),\Sigma_1(\sigma')}=
\frac{T}{2}\psi(\sigma) \delta(\sigma-\sigma') \ , \nonumber \\
& &\pb{p_\psi(\sigma),\Sigma_2(\sigma')}=-\frac{T}{2}e_1^{ \ 1}(\sigma)\delta(\sigma-\sigma') \ , 
 \quad \pb{p_\omega(\sigma),\Sigma_2(\sigma')}=
-\frac{T}{2}e_1^{ \ 0}(\sigma)\delta(\sigma-\sigma') \ , \nonumber \\ 
& &\pb{\pi^1_{ \ 0}(\sigma),\Sigma_2(\sigma')}=
-\frac{T}{2}\omega(\sigma) \delta(\sigma-\sigma') \ , \quad 
\pb{\pi^1_{ \ 1}(\sigma),\Sigma_2(\sigma')}=
-\frac{T}{2}\psi(\sigma)\delta(\sigma-\sigma') \ . \nonumber \\
& &\pb{\Sigma_1(\sigma),\Sigma_2(\sigma')}=-Tv^\mu m_\mu (\sigma')\partial_{\sigma'}
\delta(\sigma-\sigma')-Tm_\nu (\sigma)\partial_\sigma \delta(\sigma-\sigma)v^\mu(\sigma')=\nonumber \\
& &=Tv^\mu \partial_\nu m_\mu \partial_\sigma x^\nu\delta(\sigma-\sigma') \ , 
\nonumber \\
& &\pb{\Sigma_2(\sigma),\Sigma_2(\sigma)}=\pb{\Sigma_1(\sigma),\Sigma_1(\sigma')}=0
\nonumber \\
\end{eqnarray}
using the fact that
\begin{equation}
f(\sigma')\partial_\sigma\delta(\sigma-\sigma')=
f(\sigma)\partial_\sigma \delta(\sigma-\sigma)+\partial_\sigma f(\sigma)
\delta(\sigma-\sigma') \ . 
\end{equation}
We see that there is non-zero Poisson bracket between $\pi^1_{ \ 0},\pi^1_{ \ 1}$ and $\Sigma_{1,2}$
which makes analysis slightly complicated. In order to resolve
this issue let us 
 introduce 
 $\tilde{\pi}^1_{ \ 0}$ as a specific linear combinations
of primary constraints that has vanishing Poisson brackets with 
$\Sigma_1,\Sigma_2$. Explicitly, we have
\begin{equation}
\tpi^1_{ \ 0}=\pi^1_{ \ 0}-\frac{1}{e_1^{ \ 0}}\psi p_\psi-\frac{1}{e_1^{ \ 0}}\omega p_\omega +\pi^1_{ \ 1}\frac{e_1^{ \ 1}}{e_1^{ \ 0}}
\end{equation}
that obeys
\begin{eqnarray}
\pb{\tpi^1_{ \ 0},\Sigma_1}=0 \ , \quad  
\pb{\tpi^1_{ \ 0},\Sigma_2}=0 \ . 
\end{eqnarray}
In the same way we introduce $\tpi^1_{ \ 1}$ defined as
\begin{equation}
\tpi^1_{ \ 1}=\pi^1_{ \ 1}-\frac{1}{e_1^{ \ 0}}\psi p_{\omega}
\end{equation}
that clearly obeys
\begin{equation}
\pb{\tpi^1_{ \ 1},\Sigma_1}=0 \ , \quad 
\pb{\tpi^1_{ \ 1},\Sigma_2}=0 \ .
\end{equation}
In the same way we have
\begin{eqnarray}
& &\pb{\tpi^1_{ \ 0},p_\psi}\approx 0 \ , \quad \pb{\tpi^1_{ \ 0},p_\omega}\approx 0 \ , \quad \pb{\tpi^1_{ \ 1},p_\psi}\approx 0 \ , \quad 
\pb{\tpi^1_{ \ 1},p_\omega}\approx 0 \ , \nonumber \\
& & \pb{\tpi^1_{ \ 0},\pi^\alpha_{ \ a}}\approx 0 \ \quad \ , \pb{\tpi^1_{ \ 1},\pi^\alpha_{ \ a}}\approx 0 \ . 
\end{eqnarray}
Note that $\pi^0_{ \ 1}\approx 0 \ , \pi^0_{ \ 0}$ are unchanged. Then clearly
$\tpi^1_{ \ 1}\approx 0, \tpi^1_{ \ 0}\approx 0$ and $\pi^0_{ \ 0}
\approx 0, \pi^0_{ \ 1}\approx 0$ are first class constraints. 

Now we are ready to study preservation of the primary constraints. In case of $p_\omega\approx 0$ we get
\begin{eqnarray}\label{prespomega}
\partial_0 p_\omega=\pb{p_\omega,H_E}=-\frac{T}{2}e_0^{ \ 0}\tau_1-\Omega^2\frac{T}{2} e_1^{ \ 0}=0 \ , 
\nonumber \\
\end{eqnarray}
where $H_E=\int d\sigma \mH_E$. Note that (\ref{prespomega}) can
be solved for  $\Omega^2$ as
\begin{equation}
\Omega^2=-\tau_1\frac{e_0^{ \ 0}}{e_1^{ \ 0}} \ .
\end{equation}
 Further, condition of the preservation of the constraint $p_\psi\approx 0$ implies
\begin{equation}
\partial_0 p_\psi=\pb{p_\psi,H_E}=
-\frac{T}{2}(e_0^{ \ 1}\tau_1+e_0^{ \ 0}\partial_1\eta)
+\Omega^1 \frac{T}{2} e_1^{ \ 0}-\Omega^2 \frac{T}{2}e_1^{ \ 1}=0
\end{equation}
that can be solved for $\Omega^1$ as
\begin{equation}
\Omega^1=-\frac{e}{e_1^{ \ 0}}\tau_1+e_0^{ \ 0}
\partial_1 \eta \ . 
\end{equation}
Let us finally analyse conditions of preservation of constraints $\Sigma_1\approx 0$ and $\Sigma_2\approx 0$. In case of $\Sigma_1\approx 0$ we obtain
\begin{eqnarray}
& &\partial_0 \Sigma_1(\sigma)=\pb{\Sigma_1(\sigma),H_E}=
\int d\sigma' (\pb{\Sigma_1(\sigma),\mH_B(\sigma')}
+\Omega_\psi\pb{\Sigma_1(\sigma),p_\psi(\sigma')}+\nonumber \\
& &+\Omega^2\pb{\Sigma_1(\sigma),\Sigma_2(\sigma')})=0
\nonumber \\
\end{eqnarray}
which is equation for $\Omega_\psi$. In the same way requirement of the 
preservation of the constraint $\Sigma_2(\sigma)\approx 0$ implies
\begin{eqnarray}
& &\partial_0 \Sigma_2(\sigma)=\pb{\Sigma_2(\sigma),H_E}=
\int d\sigma' (\pb{\Sigma_2(\sigma),\mH_B(\sigma')}
+\Omega_\psi\pb{\Sigma_2(\sigma),p_\psi(\sigma')}+
\nonumber \\
& &+\Omega_\omega \pb{\Sigma_2(\sigma),p_\omega(\sigma')}+\Omega^2\pb{\Sigma_2(\sigma),\Sigma_1(\sigma')})=0
\nonumber \\
\end{eqnarray}
that, using the fact that we know $\Omega_1$ and $\Omega_\psi$ allows us to 
solve for $\Omega_\omega$. These results are consequence of the 
fact that $\Sigma_1,\Sigma_2$ and $p_\omega,p_\sigma$ are second class constraints.

As the final step we study the question of preservation of the constraints
\begin{equation}
\label{conpi}
\tpi^1_{ \ 0}\approx 0 \ , \quad 
\tpi^1_{ \ 1}\approx 0\  ,\quad   \pi^0_{ \ 1}\approx 0 \ , \quad   \pi^0_{ \ 0}\approx 0 \ . 
\end{equation}
First of all we use the fact that  $\theta^\alpha_{ \ a}$ has following
components
\begin{equation}
\theta^\alpha_{ \ a}=\left(\begin{array}{cc}
\theta^0_{ \ 0} & \theta^0_{ \ 1} \\
\theta^1_{ \ 0} & \theta^1_{ \ 1} \\ \end{array}\right)=\frac{1}{e}\left(\begin{array}{cc}
e_1^{ \ 1} & -e_1^{ \ 0} \\
-e_0^{ \ 1} & e_0^{ \ 0} \\ \end{array}\right)
\end{equation}
so that $\mH_B$ is equal to 
\begin{eqnarray}
& &\mH_B=-\frac{e}{2T e_1^{ \ 0}e_1^{ \ 0}}
[p_\mu h^{\mu\nu}p_\nu+2Tp_\mu h^{\mu\nu}m_\nu\partial_1\eta+T^2\partial_1\eta
m_\mu h^{\mu\nu}m_\nu\partial_1\eta]+\nonumber \\
& &+\frac{e_0^{ \ 0}}{e_1^{ \ 0}}(p_\mu \partial_1 x^\mu+p_\mu v^\mu \tau_1+T\partial_1\eta m_1+
T\partial_1 \eta m_\mu v^\mu \tau_1)+\nonumber \\
& &+\frac{T}{2}\omega e_0^{ \ 0}\tau_1
+\frac{T}{2}\psi (e_0^{ \ 1}\tau_1+e_0^{ \ 0}\partial_1\eta) \ .  \nonumber \\
\end{eqnarray}
To proceed further we use the fact that 
\begin{equation}
\pb{\pi^\alpha_{ \ a}(\sigma),e(\sigma')}=
\pb{\pi^\alpha_{ \ a}(
	\sigma),\det e_\beta^{ \ b}(\sigma')}=
-\theta^\alpha_{ \ a}e(\sigma)\delta(\sigma-\sigma') \ . 
\end{equation}
Then we start with the requirement of the preservation 
of constraint $\pi^0_{ \ 0}$ and we obtain
\begin{eqnarray}
& &\partial_0 \pi^{0}_{ \ 0}=\pb{\pi^0_{ \ 0},H_E}=
\frac{e_1^{ \ 1}}{2Te_1^{ \ 0}e_1^{ \ 0}}
[p_\mu h^{\mu\nu}p_\nu+2Tp_\mu h^{\mu\nu}m_\nu\partial_1\eta+T^2\partial_1\eta
m_\mu h^{\mu\nu}m_\nu\partial_1\eta]-\nonumber \\
& &-\frac{1}{e_1^{ \ 0}}
(p_\mu \partial_1 x^\mu+p_\mu v^\mu \tau_1+T\partial_1\eta m_1+
T\partial_1 \eta m_\mu v^\mu \tau_1)
-\frac{T}{2}\omega \tau_1-
\frac{T}{2}\psi \partial_1 \eta=
\nonumber \\
&& =\frac{e_1^{ \ 1}}{2Te_1^{ \ 0}e_1^{ \ 0}}
[p_\mu h^{\mu\nu}p_\nu+2T p_\mu h^{\mu\nu}m_\nu \partial_1\eta+2Tp_\eta \tau_1
-2T^2 m_1\tau_1+T^2\partial_1\eta m_\mu h^{\mu\nu}m_\nu
\partial_1\eta]-\nonumber \\
& &-\frac{1}{e_1^{ \ 0}}(p_\mu\partial_1 x^\mu+p_\eta \partial_1\eta)+
\Sigma_1\left(-\frac{1}{e_1^{ \ 0}}+\frac{e_1^{ \ 1}}{e_1^{ \ 0}e_1^{ \ 0}}\right)
+\frac{1}{e_1^{ \ 0}}\Sigma_2 \nonumber \\
\nonumber \\
\end{eqnarray}
using the fact that
\begin{eqnarray}
& &\psi=\frac{2}{Te_1^{ \ 0}}
(-\Sigma_1+p_\eta-Tm_1) \ , \nonumber \\
& &\omega=\frac{2}{Te_1^{ \ 0}}
(\Sigma_2+\Sigma_1\frac{e_1^{ \ 1}}{e_1^{ \ 0}}-v^\mu p_\mu
-T v^\mu m_\mu\partial_1\eta-\frac{e_1^{ \ 1}}{e_1^{ \ 0}}p_\eta+\frac{e_1^{ \ 1}}{e_1^{ \ 0}}Tm_1)
\nonumber \\
\end{eqnarray}
as follows from the definition of the primary constraints
$\Sigma_1,\Sigma_2$. 

In the same way we can proceed with the time evolution of
constraint 
 $\pi^0_{ \ 1}$ and we  get
\begin{eqnarray}
& &\partial_0 \pi^0_{ \ 1}=\pb{\pi^0_{ \ 1},H_E}=\nonumber \\
& &=-\frac{1}{2T e_1^{ \ 0}}
[p_\mu h^{\mu\nu}p_\nu+2T p_\mu h^{\mu\nu}m_\nu \partial_1 \eta+
2T p_\eta \tau_1-2T^2 m_1\tau_1+T^2\partial_1\eta
m_\mu h^{\mu\nu}m_\nu\partial_1\eta]+\frac{1}{e_1^{ \ 0}}\Sigma_1 \ . 
\nonumber \\
\end{eqnarray}
In case of $\tpi^1_{ \ 1}\approx 0$ we obtain
\begin{eqnarray}
& &\partial_0 \tpi^1_{ \ 1}=\pb{\tpi^1_{ \ 1},H_E}=\nonumber \\
& &=\frac{e_0^{ \ 0}}{2Te_1^{ \ 0} e_1^{ \ 0}}
[p_\mu h^{\mu\nu}p_\nu+2T p_\mu h^{\mu\nu}m_\nu \partial_1 \eta+
2T p_\eta \tau_1-2T^2 m_1\tau_1+
T^2
\partial_1\eta m_\mu h^{\mu\nu}m_\nu\partial_1\eta]-\frac{e_0^{ \ 0}}{e_1^{ \ 0}e_1^{ \ 0}}\Sigma_1 \ . 
\nonumber \\
\end{eqnarray}
In the same way we can proceed with 
 $\tpi^1_{ \ 0}$ and we obtain that all constraints (\ref{conpi}) 
 are preserved when we introduce two secondary constraints
\begin{eqnarray}\label{secondcon}
& &\mH_1=
p_\mu h^{\mu\nu}p_\nu+2T p_\mu h^{\mu\nu}m_\nu \partial_1 \eta+
2T p_\eta \tau_1-2T^2 m_1\tau_1+
T^2
\partial_1\eta m_\mu h^{\mu\nu}m_\nu\partial_1\eta\approx 0 \ , \nonumber \\
& &\mH_2=p_\eta\partial_1\eta+p_\mu\partial_1 x^\mu \approx 0 \ . \nonumber \\
\end{eqnarray}
Note also that using these secondary constraints the Hamiltonian density $\mH_B$ can
be written as 
\begin{eqnarray}
\mH_B=-\frac{e}{2Te_1^{ \ 0}e_1^{ \ 0}}\mH_1+
\frac{e_0^{ \ 0}}{e_1^{ \ 0}}\mH_2+
+\frac{e_0^{ \ 0}}{e_1^{ \ 0}}\tau_1(\Sigma_2+\frac{e_1^{ \ 1}}{e_1^{ \ 0}}\Sigma_1)
-\frac{\Sigma_1}{e_1^{ \ 0}}(e_0^{ \ 1}\tau_1+e_0^{\ 0}\partial_1\eta) \ . \nonumber \\
\end{eqnarray}
We see that Hamiltonian is linear combinations of constraints. As the last step we should analyse Poisson brackets between constraints $\mH_1$ and $\mH_2$. Since they contain spatial derivatives of $x^\mu$ it is convenient to introduce their smeared form multiplied by arbitrary functions $N^1,M^1$ and $N^2,M^2$. Explicitly, we have
\begin{eqnarray}
\bT^{1,2}(N^{1,2})
\equiv \int d\sigma N^{1,2}\mH_{1,2} \ , \quad  \bT^{1,2}(M^{1,2})=\int d\sigma 
M^{1,2}\mH_{1,2} \ . 
\end{eqnarray}
Then using standard Poisson brackets we 
 obtain
\begin{eqnarray}
& &\pb{\bT^1(N^1),\bT^1(M^1)}=0 \ , \nonumber \\
& &\pb{\bT_2(N^2),\bT_2(M^2)}=
\bT_2(N^2\partial_1 M^2-
M^2\partial_1 N^2) \ . \nonumber \\
\end{eqnarray}
Finally we determine Poisson bracket between generator of spatial
diffeomorphism $\bT^2(N^2)$ and $\mH_1$ and we obtain
\begin{eqnarray}
\pb{\bT^2(N^2),\mH_1(\sigma)}
=-2\partial_1 N^2 \mH_1-N^2\partial_1 \mH_1\approx 0
\nonumber \\
\end{eqnarray}
which shows that $\mH_1$ is tensor density. 
These results show that $\mH_1\approx 0 \ , \mH_2\approx 0$ are correct
form of diffeomorphism constraints which is consequence of the fact
that action for SMT string is still diffeomorphism invariant.

Finally we should analyse conditions of the preservation of constraints
$\mH_1\approx 0, \mH_2\approx 0$. 
We see that generally Poisson brackets between  $\mH_{1,2}$
and $\Sigma_1,\Sigma_2$ do not vanish. Instead we know that $\Sigma_1,\Sigma_2$ have
non-zero Poisson brackets between $p_\psi,\psi_\omega$ so that they can
be interpreted as second class constraints. 
Let us denote these second class
constraints as $\Psi^A=(p_\omega,\Sigma_1,p_\psi,\Sigma_2)$ 
with following structure of Poisson brackets
\begin{equation}
\pb{\Psi^A(\sigma),\Psi^B(\sigma')}=\triangle^{AB}(\sigma,\sigma') \ , 
\end{equation}
where
\begin{equation}\label{triangle}
\triangle^{AB}=\left(\begin{array}{cccc}
0 & \frac{1}{2}e_1^{ \ 0} & 0 & -\frac{1}{2}e_1^{\ 1} \\
-\frac{1}{2}e_1^{ \ 0} & 0 & 0 & v^\mu\partial_1 m_\mu \\
0 & 0 & 0 & -\frac{1}{2}e_1^{ \ 0} \\
\frac{1}{2}e_1^{ \ 1} &-v^\mu\partial_1 m_\mu &  \frac{1}{2}e_1^{ \ 0} & 0 \\
\end{array}
\right)T\delta(\sigma-\sigma')
\end{equation}
with inverse matrix
\begin{equation}\label{trianglein}
\triangle_{AB}=\frac{2}{T}\left(\begin{array}{cccc}
0 & -\frac{1}{e_1^{ \ 0} } & -2\frac{v^\mu\partial_1 m_\mu}{e_1^{ \ 0}e_1^{ \ 0}} & 0 \\
\frac{1}{e_1^{ \ 0}} &0 & -\frac{e_1^{ \ 1}}{e_1^{ \ 0}e_1^{ \ 0}} & 0 \\
2\frac{v^\mu\partial_1 m_\mu}{e_1^{ \ 0}e_1^{ \ 0}} & 
\frac{e_1^{ \ 1}}{e_1^{ \ 0}e_1^{ \ 0}} & 0 &\frac{1}{e_1^{ \ 0}} \\
0 & 0 &-\frac{1}{e_1^{ \ 0}} & 0 \\ \end{array}\right)\delta(\sigma-\sigma') \ . 
\end{equation}
Let us then introduce modified constraints $\tmH_i,i=1,2$ as
\begin{equation}
\tmH_i=\mH_i-\Psi^A\triangle_{AB}\pb{\Psi^B,\mH_i} \ ,
\end{equation}
where summation over $A$ includes also integration over $\sigma$  implicitly. 
Using the fact that $\pb{\mH_i,\mH_j}\approx 0$ we easily get that
\begin{equation}
\pb{\tmH_i,\tmH_j}\approx 0 \ . 
\end{equation}
Then we have
\begin{equation}
\pb{\tmH_i,\Psi^A}=\pb{\tmH_i,\Psi^A}+\pb{\Psi^A,\Psi^C}\triangle_{CB}\pb{\Psi^B,\mH_i}\approx 0
\end{equation}
and hence $\tmH_i$ have vanishing Poisson brackets with all constraints.
On the other hand since $\Psi^A$ are second class constraints that vanish
strongly in the end of the procedure we find   that $\tmH_i$ coincide 
with $\mH_i$.  Of course, this can be done on condition that we replace ordinary
Poisson brackets by Dirac brackets whose structure will be studied in the next
section.
\subsection{Symplectic Structure}
We saw above that $\Psi^A$ are second class constraints with the matrix of Poisson 
brackets given in (\ref{triangle}) and its inverse given in (\ref{trianglein}).
In order to determine Dirac brackets between canonical variables we firstly
calculate Poisson brackets between 
 canonical variables and second class constraints $\Psi^A$
\begin{eqnarray}
& &\pb{x^\mu(\sigma),\Psi^A(\sigma')}=(0, 0,0,v^\mu)\delta(\sigma-\sigma') \ , \nonumber \\ 
& &\pb{p_\mu(\sigma),\Psi_A(\sigma')}=(0,T\partial_\mu m_\nu\partial_1 x^\nu
\delta(\sigma-\sigma')+Tm_\mu(\sigma')\partial_{\sigma'}\delta(\sigma-\sigma'), \nonumber \\  
& & 0 ,
-\partial_\mu v^\nu p_\nu\delta(\sigma-\sigma')-T\partial_\mu (v^\nu m_\nu)\partial_1\eta
\delta(\sigma-\sigma')) \ , \nonumber \\
& &\pb{\eta(\sigma),\Psi^A(\sigma')}=(0,0,0,-Tv^\mu m_\mu(\sigma')\partial_{\sigma'}
\delta(\sigma-\sigma')) \ , \nonumber \\
& &\pb{p_\eta(\sigma),\Psi^A(\sigma)}=
(0,\delta(\sigma-\sigma'),0,0) \ .  \nonumber \\
\end{eqnarray}
Then we find following form of Dirac brackets between canonical variables
\begin{eqnarray}
& &\pb{\eta(\sigma),p_\eta(\sigma')}_D=
\pb{\eta(\sigma),p_\eta(\sigma')}-\nonumber \\
& &-\int d\sigma_1 d\sigma_2
\pb{\eta(\sigma),\Psi^A(\sigma_1)}\triangle_{AB}(\sigma_1,\sigma_2)
\pb{\Psi^B(\sigma_2),p_\eta(\sigma')}=\delta(\sigma-\sigma') \ , \nonumber \\
& &\pb{\eta(\sigma),\eta(\sigma')}_D=-
\int d\sigma_1 d\sigma_2
\pb{\eta(\sigma),\Psi^A(\sigma_1)}\triangle_{AB}(\sigma_1,\sigma_2)
\pb{\Psi^B(\sigma_2),\eta(\sigma')}=0 \ , \nonumber \\
&  &\pb{p_\eta(\sigma),p_\eta(\sigma')}_D=-
\int d\sigma_1 d\sigma_2
\pb{p_\eta(\sigma),\Psi^A(\sigma_1)}\triangle_{AB}(\sigma_1,\sigma_2)
\pb{\Psi^B(\sigma_2),p_\eta(\sigma')}=0 
\nonumber \\
& &\pb{x^\mu(\sigma),p_\nu(\sigma')}_D=
\pb{x^\mu(\sigma),p_\nu(\sigma')}-\nonumber \\
& &-\int d\sigma_1 d\sigma_2
\pb{x^\mu(\sigma),\Psi^A(\sigma_1)}\triangle_{AB}(\sigma_1,\sigma_2)
\pb{\Psi^B(\sigma_2),p_\nu(\sigma')}=\delta^\mu_\nu\delta(\sigma-\sigma')  \ , 
\nonumber \\
& &\pb{x^\mu(\sigma),x^\nu(\sigma')}_D=
-\int d\sigma_1 d\sigma_2
\pb{x^\mu(\sigma),\Psi^A(\sigma_1)}\triangle_{AB}(\sigma_1,\sigma_2)
\pb{\Psi^B(\sigma_2),x^\nu(\sigma')}0 \ , 
\nonumber \\
& &\pb{p_\mu(\sigma),p_\nu(\sigma')}_D=
-\int d\sigma_1 d\sigma_2
\pb{p_\mu(\sigma),\Psi^A(\sigma_1)}\triangle_{AB}(\sigma_1,\sigma_2)
\pb{\Psi^B(\sigma_2),p_\nu(\sigma')}=0 \ . 
\nonumber \\
\end{eqnarray}
Finally we determine mixed Dirac brackets 
\begin{eqnarray}
& &\pb{x^\mu(\sigma),\eta(\sigma')}_D=
-\int d\sigma_1 d\sigma_2
\pb{x^\mu(\sigma),\Psi^A(\sigma_1)}\triangle_{AB}(\sigma_1,\sigma_2)
\pb{\Psi^B(\sigma_2),\eta(\sigma')}=0  \ , 
\nonumber \\
& &\pb{x^\mu(\sigma),p_\eta
		(\sigma')}_D=
-\int d\sigma_1 d\sigma_2
\pb{x^\mu(\sigma),\Psi^A(\sigma_1)}\triangle_{AB}(\sigma_1,\sigma_2)
\pb{\Psi^B(\sigma_2),p_\eta(\sigma')}=0 \ , 
\nonumber \\
& &\pb{p_\mu(\sigma),\eta(\sigma')}_D=
-\int d\sigma_1 d\sigma_2
\pb{p_\mu(\sigma),\Psi^A(\sigma_1)}\triangle_{AB}(\sigma_1,\sigma_2)
\pb{\Psi^B(\sigma_2),\eta(\sigma')}=0 \ , 
\nonumber \\
& &\pb{p_\mu(\sigma),p_\eta(\sigma')}_D=
-\int d\sigma_1 d\sigma_2
\pb{p_\mu(\sigma),\Psi^A(\sigma_1)}\triangle_{AB}(\sigma_1,\sigma_2)
\pb{\Psi^B(\sigma_2),p_\eta(\sigma')}=0 \ . 
\nonumber \\
\end{eqnarray}
These results show that Dirac brackets between $p_\mu,x^\mu,p_\eta,\eta$ have
the same form as Poisson brackets. In the next section 
we consider situation when $\theta^0_{ \ 1}=0$.

\section{Singular Case}\label{third}
Canonical analysis performed in previous section was valid on condition that   $\theta^0_{ \ 1}\neq 0$ or equivalently
 $e_1^{ \ 0}\neq 0$. However spatial gauge that was 
imposed in \cite{Harmark:2018cdl,Harmark:2020vll} is valid on condition 
when  $e_1^{ \ 0}=0$. In other words this 
gauge fixing cannot be reached in previous analysis and deserves separate treatment. We call this case as singular  since, as we will see below, it will not be possible
to express time derivative of $x^\mu$ as function of canonical variables.  

To see this explicitly  we start with the action (\ref{actnonnon}) from which 
we determine following conjugate momenta
\begin{eqnarray}
& &\pi^\alpha_{ \ b}=\frac{\partial \mL}{\partial \partial_0 e_\alpha^{ \ a}}\approx 0 
\ , \quad p_\psi \approx 0 \ , \quad \ , \pi_{\omega}\approx 0 \ , \quad 
p_\eta=T m_1 \ , 
\nonumber \\
& & p_\mu=\frac{\partial \mL}{\partial \partial_0 x^\mu}=
-Tm_\mu\partial_1 \eta +\frac{T}{2}\psi e_1^{ \ 1}\tau_\mu\nonumber \\
\nonumber \\
\end{eqnarray}
that implies an existence of  primary constraints 
\begin{eqnarray}
\Sigma_1\equiv p_\eta-T m_1  \ , \quad 
\Sigma_\mu^{2} \equiv p_\mu+Tm_\mu\partial_1\eta-\frac{T}{2}\psi e_1^{ \ 1}\tau_\mu\approx 0 \ . 
\nonumber \\
\end{eqnarray}
For further purposes we introduce following linear combination 
of constraints that we denote as $\mH_2$:
\begin{equation}\label{mH2}
\mH_2\equiv\partial_1 x^\mu \Sigma_\mu^2+\Sigma_1\partial_1\eta=p_\mu \partial_1 x^\mu+
p_\eta \partial_1\eta -\frac{T}{2}\psi e_1^{ \ 1}\tau_1 \approx 0
\end{equation}
that will be useful below. 

As the next step we determine bare Hamiltonian density in the form 
\begin{eqnarray}
\mH_B=p_\mu\partial_0 x^\mu+p_\eta \partial_0 \eta-\mL=\frac{T}{2}e \theta^1_{ \ 1}\theta^1_{ \ 1}h_{11}+\frac{T}{2}\omega e_0^{ \ 0}\tau_1
+\frac{T}{2}\psi (e_0^{ \ 0}\partial_1\eta+e_0^{ \ 1}\tau_1) \ . 
\nonumber \\
\end{eqnarray}
Let us now proceed to the  analysis of  preservation of primary constraints. We introduce extended Hamiltonian as
\begin{equation}
H_E=\int d\sigma (\mH_B+\Omega^1 \Sigma_1+\Omega^\mu_2 \Sigma_\mu^2+
v^\psi p_\psi+v^\omega p_\omega) \ .
\end{equation}
We observe that we can always write $\Omega^1=\tilde{\Omega}^1\partial_1 \eta \Sigma_1$
so that when we use (\ref{mH2})  we can express $\partial_1\eta\Sigma_1$ with the help of $\mH_2$ and hence extended Hamiltonian density $\mH_E$ can be written in the form  
\begin{eqnarray}
& &\mH_E=T\frac{e_0^{ \ 0}e_0^{ \ 0}}{2e}h_{11}+\frac{T}{2}\omega e_0^{ \ 0}\tau_1+
\frac{T}{2}\psi(e_0^{ \ 0}\partial_1\eta+e_0^{ \ 1}\tau_1)+
\nonumber \\
& &+v^\psi p_\psi+v^\omega p_\omega +\tilde{\Omega}^1\mH_2+\tilde{\Omega}^\mu_2\Sigma^2_\mu \ , 
\nonumber \\
\end{eqnarray}
where we introduced $\tilde{\Omega}^\mu_2$ as
$\tilde{\Omega}^\mu_2=\Omega^\mu_2-
\tilde{\Omega}\partial_1 x^\mu$. Then in what follows we will omit tilde on $\Omega's$.

 Now we are ready to analyse requirement of the preservation of all constraints. 
 In case of $p_\omega$ we get
\begin{equation}
\partial_0 p_\omega=\pb{p_\omega,H_E}=-\frac{T}{2}e_0^{ \ 0}\tau_1\equiv -\frac{T}{2}e_0^{ \ 0}\Sigma_\omega^{II}\approx 0 \ ,  
\end{equation}
where $\Sigma^{II}_\omega=\tau_1 \approx 0$ is new secondary constraint. Generally 
this constraint would imply $\partial_1 x^\mu=0$ however this is very strong condition. We should rather presume that the background has non-zero component $\tau_0$ only so that this constraint is equal to $\Sigma^{II}_{\omega}\equiv \partial_1 x^0\approx 0$. As a consequence  $\mH_2$ is standard spatial diffeomorphism constraint which is the first class constraint. 

Now using the fact that $\tau_0\neq 0 \ ,\tau_i=0$ we have 
\begin{equation}
\Sigma_0^2=p_0+T m_0\partial_1\eta-\frac{T}{2}\psi e_1^{ \ 1}\tau_0 \ , \quad 
\Sigma_i^2=p_i+Tm_i\partial_1\eta \ . 
\end{equation}
For further purposes we calculate Poisson brackets between primary constraints
\begin{eqnarray}
& &\pb{p_\psi(\sigma),\Sigma_0^2(\sigma')}=\frac{T}{2}e_1^{ 1}\tau_0\delta(\sigma-\sigma') \ , \nonumber \\
& &\pb{\Sigma^2_i(\sigma),\Sigma^2_j(\sigma')}=-T(\partial_i m_j-\partial_j m_i)
\partial_1 \eta \delta(\sigma-\sigma')\equiv -\mF_{ij}\delta(\sigma-\sigma') \ .
\nonumber \\
\end{eqnarray}
In the same way we denote Poisson bracket between $\Sigma_0^2$ and $\Sigma_i^2$ as
\begin{equation}
\pb{\Sigma_i^2(\sigma),\Sigma^2_0(\sigma')}=-\mF_{i0}\delta(\sigma-\sigma') \ . 
\end{equation}
Let us now study the requirement of the preservation of constraint $p_\psi$
\begin{equation}
\partial_0 p_\psi=\pb{p_\psi,H_E}=
-\frac{T}{2}e_0^{ \ 0}\partial_1\eta+\Omega^0_2 \frac{T}{2}e_1^{ \ 1}\tau_0=0
\end{equation}
that has solution
\begin{equation}
\Omega^0_2=\frac{e_0^{ \ 0}}{\tau_0 e_1^{ \ 1}}\partial_1\eta \ . 
\end{equation}
In other words, $\Sigma_0^2\approx 0, p_\psi\approx 0$ are second class constraints
that can be explicitly solved for $p_\psi$ and $\psi$. We return to this problem below. Instead we focus on the  time evolution of constraint $\Sigma_0^2\approx 0$ that has the form 
\begin{equation}
\partial_0 \Sigma_0^2=\pb{\Sigma_0^2,H_E}=
\int d\sigma \left(\pb{\Sigma_0^2,\mH_E}-\frac{T}{2}e_1^{ \ 1}\tau_0\delta(\sigma-\sigma')
v_\psi +\mF_{0i}\Omega^i_2\right)=0
\end{equation}
which can be solved  for $v_\psi$. Finally, the requirement of the preservation of constraints $\Sigma_i^2\approx 0$ has the form
\begin{eqnarray}
\partial_0 \Sigma_i^2=\pb{\Sigma_i^2,H_E}=
\int d\sigma' (\pb{\Sigma_i^2,\mH_B(\sigma')}
-\mF_{i0}\delta(\sigma-\sigma')\Omega_0^2+\mF_{ij}\delta(\sigma-\sigma')\Omega^j_2)=0 \ .  \nonumber \\
\end{eqnarray}
Since $\mF_{ij}$ is non-singular by definition we can solve the equation above for $\Omega^2_i$.

Let us  analyse requirement of the preservation of constraints $\pi^\alpha_{ \ a}$. Following analysis presented in section (\ref{second}) we replace $\pi^1_{ \ 1}$ with $\tpi^1_{ \ 1}$ defined as 
\begin{equation}
\tpi^1_{ \ 1}=\pi^1_{ \ 1}-\frac{\psi}{e_1^{ \ 1}}p_\psi
\end{equation}
that has vanishing Poisson bracket with $\Sigma_0^2\approx 0$. Further, requirement  of the  preservation of $\pi^0_{ \ 0}$ has the form
\begin{eqnarray}\label{eqpioo}
\partial_0 \pi^0_{ \ 0}=\pb{\pi^0_{ \ 0},H_E}
=-\frac{1}{e_1^{ \ 1}}\left[\frac{T}{2}h_{11}+\frac{1}{\tau_0}
(p_0-Tm_0\partial_1\eta)\partial_1
\eta\right]+\frac{1}{e_1^{ \ 1}\tau_0}\Sigma_0^2\approx 0
\nonumber \\
\end{eqnarray}
using the fact that
\begin{equation}
\frac{T}{2}\psi=\frac{1}{e_1^{ \ 1}\tau_0}(p_0-Tm_0\partial_1\eta-\Sigma_0^2) \ 
\end{equation}
and also that $e$ is equal to $e=\det e_\alpha^{ \ a}=e_0^{\ 0}e_1^{ \ 1}$.
We see that in order to obey equation (\ref{eqpioo}) we should introduce
secondary constraint $\mH_1$ defined as 
\begin{equation}
\mH_{1}=\frac{T}{2}h_{11}+\frac{1}{\tau_0}(p_0-Tm_0\partial_1\eta)\partial_1\eta\approx 0 \ . 
\end{equation}
On the other hand requirement of the preservation of the
constraint $\tilde{\pi}^1_{ \ 1}\approx 0$ gives
\begin{eqnarray}
\partial_0\tilde{\pi}^1_{ \ 1}=\pb{\tilde{\pi}^1_{ \ 1},H_E}=
\frac{e_0^{ \ 0}}{(e_1^{ \ 1})^2}\mH_{1}-\frac{1}{\tau_0 (e_1^{ \ 1})^2}\Sigma_0^2
\approx 0 \ . \nonumber \\
\end{eqnarray}
 Clearly 
\begin{equation}
\pb{\mH_{1}(\sigma),\mH_{1}(\sigma')}=0 \ , \quad \pb{\mH_{1}(\sigma),
\mH_{2}(\sigma')}\approx 0 \ , \quad \pb{\mH_{2}(\sigma),\mH_{2}(\sigma')}\approx 0
\end{equation}
and hence they are the first class constraints reflecting invariance of the world-sheet
theory under reparameterization. 

\subsection{Symplectic structure}
In this section we study symplectic structure of the theory studied in previous section. For simplicity of our analysis
we will consider partial fixed theory with fixed spatial diffeomorphism constraint
$\mH_{2}\approx 0$. This can be done by introducing gauge fixing function 
\begin{equation}
\mG:\eta-\sigma\approx 0
\end{equation}
Since $\pb{\mG(\sigma),\mH_{2}(\sigma')}=\delta(\sigma-\sigma')$, $\mH_{2}$
and $\mG$ are second class constraints that strongly vanish. From $\mH_{2}=0$ we
express $p_\eta$ as
\begin{equation}
p_\eta=-p_\mu\partial_1 x^\mu \ . 
\end{equation}
Further, as  we argued in previous section,  we have second class constraints 
$\Psi^A=(p_\psi,\Sigma^2_0,\Sigma^2_i)$ with following matrix of Poisson 
brackets
\begin{equation}
\pb{\Psi^A(\sigma),\Psi^B(\sigma')}=
\left(\begin{array}{ccc}
0 & \frac{T}{2}e_1^{ \ 1}\tau_0 & 0 \\
-\frac{T}{2}e_1^{ \ 1}\tau_0& 0 & \mF_{0j} \\ 
0 & -\mF_{i0} & -\mF_{ij} \\ \end{array}\right)\delta(\sigma-\sigma') \ . 
\end{equation}
For simplicity we will presume that $\mF_{0j}=0$. Then the matrix inverse to $\triangle^{AB}$ is equal to
\begin{equation}
\triangle_{AB}=\left(\begin{array}{ccc}
0 & -\frac{2}{Te_1^{ \ 1}\tau_0} & 0 \\
\frac{2}{Te_1^{ \ 1}\tau_0} &  0 & 0\\
0 & 0& -\mF^{ij} \\ \end{array}\right) \ , 
\end{equation}
where $\mF^{ij}$ is matrix inverse to $\mF_{ij}$. Further, we have Poisson 
brackets
\begin{eqnarray}
\pb{x^i(\sigma),\Psi^A(\sigma')}=(0,0,\delta^i_j)\delta(\sigma-\sigma')
\nonumber \\
\end{eqnarray}
and hence
\begin{eqnarray}
& &\pb{x^i(\sigma),x^j(\sigma')}_D=\nonumber \\
& &=-\int d\sigma_1 d\sigma_2
\pb{x^i(\sigma),\Psi^A(\sigma_1)}\triangle_{AB}(\sigma_1,\sigma_2)
\pb{\Psi^B(\sigma_2),x^j(\sigma')}=-\mF^{ij}\delta(\sigma-\sigma') \ . 
\nonumber \\
\end{eqnarray}
We see that there is non-trivial symplectic structure which is in agreement 
with the observation presented 
in \cite{Harmark:2020vll}.  Then the equation of motion for $x^i$ have the form
\begin{equation}
\partial_0 x^i=\pb{x^i,H}_D=\mF^{ik}\partial_1[\lambda h_{kl}\partial_1 x^l]-
\lambda \mF^{ik}\partial_k h_{mn}\partial_1 x^m\partial_1 x^n \ ,
\end{equation}
where we used the fact that the   Hamiltonian 
is equal to 
\begin{equation}
\mH=\lambda \mH_{1}  \ , \quad 
\mH_{1}=\frac{T}{2}h_{ij}\partial_1 x^i
\partial_1 x^j+\frac{1}{\tau_0}(p_0-Tm_0) \ , 
\end{equation}
where $\lambda$ is Lagrange multiplier and where $m_0$ and $\tau_0$ do not 
depend on $x^i$. 

To conclude, we derived symplectic structure for SMT string in the gauge when $e_1^{ \ 0}=0$ and we showed that it is non-trivial and depend on the field $m_\mu$.


%

\acknowledgments{This  work  was
	supported by the Grant Agency of the Czech Republic under the grant
	GA20-04800S. }



\begin{thebibliography}{20}
\bibitem{Maldacena:1997re}
J.~M.~Maldacena,
\emph{``The Large N limit of superconformal field theories and supergravity,''}
Int. J. Theor. Phys. \textbf{38} (1999), 1113-1133
doi:10.1023/A:1026654312961
[arXiv:hep-th/9711200 [hep-th]].
	
	\bibitem{Harmark:2014mpa}
	T.~Harmark and M.~Orselli,
\emph{``Spin Matrix Theory: A quantum mechanical model of the AdS/CFT correspondence,''}
	JHEP \textbf{11} (2014), 134
	doi:10.1007/JHEP11(2014)134
	[arXiv:1409.4417 [hep-th]].




\bibitem{Harmark:2017rpg}
T.~Harmark, J.~Hartong and N.~A.~Obers,
\emph{``Nonrelativistic strings and limits of the AdS/CFT correspondence,''}
Phys. Rev. D \textbf{96} (2017) no.8, 086019
doi:10.1103/PhysRevD.96.086019
[arXiv:1705.03535 [hep-th]].


\bibitem{Harmark:2018cdl}
T.~Harmark, J.~Hartong, L.~Menculini, N.~A.~Obers and Z.~Yan,
\emph{``Strings with Non-Relativistic Conformal Symmetry and Limits of the AdS/CFT Correspondence,''}
JHEP \textbf{11} (2018), 190
doi:10.1007/JHEP11(2018)190
[arXiv:1810.05560 [hep-th]].

\bibitem{Harmark:2019upf}
T.~Harmark, J.~Hartong, L.~Menculini, N.~A.~Obers and G.~Oling,
\emph{``Relating non-relativistic string theories,''}
JHEP \textbf{11} (2019), 071
doi:10.1007/JHEP11(2019)071
[arXiv:1907.01663 [hep-th]].


\bibitem{Harmark:2018cdl}
T.~Harmark, J.~Hartong, L.~Menculini, N.~A.~Obers and Z.~Yan,
\emph{``Strings with Non-Relativistic Conformal Symmetry and Limits of the AdS/CFT Correspondence,''}
JHEP \textbf{11} (2018), 190
doi:10.1007/JHEP11(2018)190
[arXiv:1810.05560 [hep-th]].
	
\bibitem{Harmark:2020vll}
T.~Harmark, J.~Hartong, N.~A.~Obers and G.~Oling,
\emph{``Spin Matrix Theory
	 String Backgrounds and Penrose Limits of AdS/CFT,''}
[arXiv:2011.02539 [hep-th]].


\bibitem{Andringa:2012uz}
R.~Andringa, E.~Bergshoeff, J.~Gomis and M.~de Roo,
\emph{``'Stringy' Newton-Cartan Gravity,''}
Class. Quant. Grav. \textbf{29} (2012), 235020
doi:10.1088/0264-9381/29/23/235020
[arXiv:1206.5176 [hep-th]].


\bibitem{Gomis:2020izd}
J.~Gomis, Z.~Yan and M.~Yu,
\emph{``T-Duality in Nonrelativistic Open String Theory,''}
[arXiv:2008.05493 [hep-th]].

\bibitem{Gomis:2020fui}
J.~Gomis, Z.~Yan and M.~Yu,
\emph{``Nonrelativistic Open String and Yang-Mills Theory,''}
[arXiv:2007.01886 [hep-th]].

\bibitem{Kluson:2020aoq}
J.~Kluso\v{n},
\emph{``Unstable D-brane in Torsional Newton-Cartan Background,''}
JHEP \textbf{09} (2020), 191
doi:10.1007/JHEP09(2020)191
[arXiv:2001.11543 [hep-th]].


\bibitem{Hansen:2020pqs}
D.~Hansen, J.~Hartong and N.~A.~Obers,
\emph{``Non-Relativistic Gravity and its Coupling to Matter,''}
JHEP \textbf{06} (2020), 145
doi:10.1007/JHEP06(2020)145
[arXiv:2001.10277 [gr-qc]].

\bibitem{Yan:2019xsf}
Z.~Yan and M.~Yu,
\emph{``Background Field Method for Nonlinear Sigma Models in Nonrelativistic String Theory,''}
JHEP \textbf{03} (2020), 181
doi:10.1007/JHEP03(2020)181
[arXiv:1912.03181 [hep-th]].

\bibitem{Kluson:2019xuo}
J.~Kluso\v{n},
\emph{``T-duality 
	of Non-Relativistic String in Torsional Newton-Cartan Background,''}
JHEP \textbf{05} (2020), 024
doi:10.1007/JHEP05(2020)024
[arXiv:1909.13508 [hep-th]].

\bibitem{Bergshoeff:2019pij}
E.~A.~Bergshoeff, J.~Gomis, J.~Rosseel, C.~\c{S}im\c{s}ek and Z.~Yan,
\emph{``String Theory and String Newton-Cartan Geometry,''}
J. Phys. A \textbf{53} (2020) no.1, 014001
doi:10.1088/1751-8121/ab56e9
[arXiv:1907.10668 [hep-th]].

\bibitem{Gallegos:2019icg}
A.~D.~Gallegos, U.~G\"ursoy and N.~Zinnato,
\emph{``Torsional Newton Cartan gravity from non-relativistic strings,''}
JHEP \textbf{09} (2020), 172
doi:10.1007/JHEP09(2020)172
[arXiv:1906.01607 [hep-th]].

\bibitem{Gomis:2019zyu}
J.~Gomis, J.~Oh and Z.~Yan,
\emph{``Nonrelativistic String Theory in Background Fields,''}
JHEP \textbf{10} (2019), 101
doi:10.1007/JHEP10(2019)101
[arXiv:1905.07315 [hep-th]].





\bibitem{Kluson:2019ifd}
J.~Kluso\v{n},
\emph{``$(m,n)$-String and D1-Brane in Stringy Newton-Cartan Background,''}
JHEP \textbf{04} (2019), 163
doi:10.1007/JHEP04(2019)163
[arXiv:1901.11292 [hep-th]].

\bibitem{Kluson:2018vfd}
J.~Kluso\v{n},
\emph{``Note About T-duality of Non-Relativistic String,''}
JHEP \textbf{08} (2019), 074
doi:10.1007/JHEP08(2019)074
[arXiv:1811.12658 [hep-th]].

\bibitem{Kluson:2018grx}
J.~Kluso\v{n},
\emph{``Nonrelativistic String Theory Sigma Model and Its Canonical Formulation,''}
Eur. Phys. J. C \textbf{79} (2019) no.2, 108
doi:10.1140/epjc/s10052-019-6623-9
[arXiv:1809.10411 [hep-th]].

\bibitem{Bergshoeff:2018yvt}
E.~Bergshoeff, J.~Gomis and Z.~Yan,
\emph{``Nonrelativistic String Theory and T-Duality,''}
JHEP \textbf{11} (2018), 133
doi:10.1007/JHEP11(2018)133
[arXiv:1806.06071 [hep-th]].

\bibitem{Kluson:2018egd}
J.~Kluso\v{n},
\emph{``Remark About Non-Relativistic String in Newton-Cartan Background and Null Reduction,''}
JHEP \textbf{05} (2018), 041
doi:10.1007/JHEP05(2018)041
[arXiv:1803.07336 [hep-th]].

\bibitem{Cartan:1923zea}
E.~Cartan,
\emph{``Sur les variétés à connexion affine et la théorie de la relativité généralisée. (première partie),''}
Annales Sci.\ Ecole Norm.\ Sup.\  {\bf 40} (1923) 325.


\end{thebibliography}
\end{document}